\documentclass[aps,prb,preprint,showpacs,superscriptaddress]{revtex4-1}
\usepackage{amsmath}
\usepackage{graphicx}
\graphicspath{{pict/}{./}}
\usepackage{bm}%
\newcommand\pictc[5]{\begin{figure}
                       \centerline{\vspace{-1mm}
\includegraphics[width=#1\columnwidth,height=0.7\textheight,keepaspectratio]{#3}}
                       \protect\caption{\protect\label{#4} #5}\vspace{-3mm}
                    \end{figure}            }

\newcommand\pict[4][1]{\pictc{#1}{!tb}{#2}{#3}{#4}}
\newcommand\rpict[1]{\ref{#1}}

\newcounter{Fig}

\newcommand{\be}{\begin{equation}}
\newcommand{\ee}{\end{equation}}

\begin{document}
\title{Toroidal dipole induced transparency in core-shell nanoparticles}
\author{Wei Liu}
\email{Corresponding author. Email: wei.liu.pku@gmail.com}
\affiliation{College of Optoelectronic Science and Engineering, National University of Defense
Technology, Changsha, Hunan 410073, China}
\affiliation{Nonlinear Physics Centre, Australian National University,
Canberra, ACT 0200, Australia}
\author{Jianfa Zhang}
\affiliation{College of Optoelectronic Science and Engineering, National University of Defense
Technology, Changsha, Hunan 410073, China}
\author{Andrey E. Miroshnichenko}
\affiliation{Nonlinear Physics Centre, Australian National University,
Canberra, ACT 0200, Australia}


\begin{abstract}
We investigate the scattering properties of spherical nanoparticles by employing a Cartesian multipole expansion method which has incorporated radiating toroidal multipoles. It is shown that toroidal dipoles, which are negligible under long-wavelength approximations, can be excited within high-permittivity dielectric nanoparticles and significantly influence the scattering profile in the optical regime. We further reveal that the scattering transparencies of core-shell plasmonic nanoparticles can be classified into two categories: i) the trivial transparency with no effective multipole excitations within the particle, and ii) the non-trivial one induced by the destructive interferences of induced electric and toroidal multipoles. The incorporation of toroidal moments offers new insights into the study into nanoparticle scattering in both the near- and far-fields, which may shed new light to many related applications, such as biosensing, nanoantennas, photovoltaic devices and so on.
\end{abstract}
\maketitle

\section{Introduction}

To obtain the scattering properties of radiation sources of arbitrary charge-current distributions  the conventional Cartesian multipole expansion method is usually employed~\cite{jackson1962classical}.  However, recently it was demonstrated that this approach is incomplete, and, for example, the toroidal multipoles have not been taken into account~\cite{Dubovik1974,Dubovik1990_PR,Afanasiev1992,Afanasiev1995,Radescu2002_PRE}. Being physically different from the electric (induced by the separation of oscillating positive and negative charges) and magnetic (induced by the enclosed circulation of electric currents) multipoles, toroidal multipoles manifest themselves as poloidal currents which flow on the surface of a torus along its meridians~\cite{Dubovik1974,Dubovik1990_PR,Afanasiev1992,Afanasiev1995,Radescu2002_PRE}. Despite the differences in terms of charge-current distributions, in the far field however the scattering patterns and parity properties of toroidal multipoles and their electric  counterparts are identical even though they correspond to inequivalent vector potentials which cannot be gauge-transformed into each other~\cite{Afanasiev1995}. Moreover, the toroidal multipoles interact only with the time derivatives of the incident fields, and consequently under most circumstances are relatively weak and, consequently, overshadowed by the much stronger electric and/or magnetic multipoles~\cite{Dubovik1990_PR,Afanasiev1995}. As a result of all the features mentioned above, only recently the electromagnetic toroidal multipoles (or more specifically toroidal dipole) and related effects have been observed on carefully designed metamaterial and plasmonic platforms~\cite{Kaelberer2010_Science,dong2012toroidal,Huang2012_OE,Ogut2012_NL,Fedotov2013_SR,Fan2013_lowloss,dong2013all,savinov2014toroidal,Basharin2014_arXiv,miroshnichenko2014seeing}.

In the currently flourishing fields of plasmonics and metamaterials, the investigation into Mie scattering of nanoparticles is of fundamental importance~\cite{Alu2005_PRE,Wheeler2006_PRB,Maier2007,Lukyanchuk2011_NM,Miroshnichenko2010_RMP,Paniague2011_NJP,Zheludev2012_NM,Liu2012_ACSNANO,Liu2014_CPB,Liu2014_ultradirectional}. One of the standard approaches in Mie scattering related studies is to decompose the fields of a spherical particle  into vector spherical electric and magnetic multipoles~\cite{Bohren1983_book,Doyle1989_optical,Wheeler2006_PRB,Paniague2011_NJP}. Since vector spherical harmonics form a full basis, meaning that any vectorial field can be uniquely decomposed, the presence toroidal modes have also been overlooked. Even though the toroidal multipoles are highly related to both the local near-field configurations and the far-field scattering patterns, and thus may play a significant role in various applications including sensing~\cite{Kabashin2009_NM}, nanoantennas~\cite{Novotny2011_NP}, and photovoltaic devices~\cite{Atwater2010_NM}, investigations into their excitations within individual Mie scattering particles are still rare~\cite{miroshnichenko2014seeing}. Moreover, considering that the electromagnetic light-matter interactions involving toroidal multipoles can possibly merge with the emerging fields of topological~\cite{Lu2014_topological} and two-dimensional photonics~\cite{Xia2014_2D}, it is essential to restudy the scattering of nanoparticles from the new perspectives of toroidal multipoles.

In this paper we revisit the seminal problem of Mie scattering of spherical nanoparticles based on the Cartesian multipole expansion method with the included toroidal components. We demonstrate that although within metallic spheres the toroidal dipole (TD) is negligible compared to electric dipole (ED), it can be excited in high-permittivity dielectric nanoparticles and strongly affect  the  scattering properties in the optical regime. We further study the hybrid metal-dielectric core-shell nanoparticles, and identify two different mechanisms of the scattering transparencies of those particles: i) the trivial transparency without effective multipole excitation inside and ii) the {\em non-trivial} one caused by scattering cancellation of electric and toroidal dipoles.  The non-trivial transparency in the optical regime we have achieved in individual core-shell nanoparticles corresponds exactly to the recently demonstrated non-trivial non-radiating charge-current distributions~\cite{Fedotov2013_SR,Basharin2014_arXiv,miroshnichenko2014seeing}.

\section{Dipole expansion for scattering particles including toroidal components}

The scattered fields of spherical particles with incident plane waves can be fully characterized by Mie scattering coefficients $a_m$ and $b_m$ (see the appendix for more information), which conventionally have been associated directly with the magnitudes of the electric and magnetic moments of order $m$ excited within the particles~\cite{Bohren1983_book,Doyle1989_optical,Wheeler2006_PRB}. For example, the electric dipole (ED) moment can be expressed as:
\begin{equation}
\label{conventioanl_dipole}
\textbf{P}(a_1)= \varepsilon _0 {{6\pi ia_1 } \over {k^3 }}\textbf{E}_0,
\end{equation}
where $\varepsilon _0$ is the vacuum permittivity, $k$ the angular wave number in the background (it is vacuum in this paper), and $\textbf{E}_0$ is the electric field of the incident wave.
Equation (\ref{conventioanl_dipole}) has been established  by the comparison of the scattered field associated with $a_1$ and that of a typical electric dipolar moment~\cite{Doyle1989_optical,Wheeler2006_PRB}, and thus the near-field charge-current distribution of the scattering particle has not been considered.  Since the far-field scattering of an ED and a TD are indistinguishable, the fields associated with $a_1$ in fact includes {\em both} contributions of ED and TD, while erroneously the scattering has usually been attributed to the ED only, with the TD being overlooked.

An alternative method to conventional spherical vector modes expansion of Mie scattering particles is Cartesian multipoles which allows to explicitly take into account toroidal moments~\cite{Radescu2002_PRE,Kaelberer2010_Science,Fedotov2013_SR,miroshnichenko2014seeing}. In Mie theory, the near-field distributions (in terms of both \textbf{E} and \textbf{H}) can be analytically expressed~\cite{Bohren1983_book} and based on the charge conservation relation the current (displacement current) distribution is [the $\rm exp(-i\omega t+i\textbf{k}\cdot\textbf{r})$ notation has been adopted for electromagnetic waves]:
\begin{equation}
\label{current}
\textbf{J}(\textbf{r}) = -i\omega \varepsilon _0 [\varepsilon(\textbf{r})-1]\textbf{E}(\textbf{r}),
\end{equation}
where $\varepsilon(\textbf{r})$ is the relative permittivity and $\omega$ is the angular frequency of the incident wave. With the current distribution, the ED and TD moments can be directly obtained~\cite{Radescu2002_PRE,Kaelberer2010_Science,Fedotov2013_SR}:
\begin{equation}
\label{ED_TD}
\textbf{P}= {1 \over { - i\omega }}\int d^3r\textbf{J}(\textbf{r}), ~~\textbf{T}= {1 \over {10c}}\int d^3r[(\textbf{r} \cdot \textbf{J}(\textbf{r}))\textbf{r} - 2r^2 \textbf{J}],
\end{equation}
where $c$ is the speed of light. It is worth mentioning here that throughout this work we use $\textbf{P}(a_1)$  and  $\textbf{P}$ to indicate the ED moment deduced from far-field scattering [see Eq.~(\ref{conventioanl_dipole})] and that obtained through current integration [see Eq.~(\ref{ED_TD})], respectively. With the presence of only ED and TD when other multipoles are negligible, the far-field radiation can be expressed as (in terms of the electric field)~\cite{Radescu2002_PRE,Chen2011_NP531}:
\begin{equation}
\label{Scattering}
\textbf{E}_{\rm rad}  = {{k^2 \exp (ikr)} \over {4\pi r\varepsilon _0 }}\textbf{n} \times (\textbf{P} + ik\textbf{T}) \times \textbf{n},
\end{equation}
where $\textbf{n}$ is the unit vector along $\textbf{r}$ . Equation~(\ref{Scattering}) confirms that ED and TD have an identical far-field scattering pattern despite an extra scaling factor $ik$ of TD. Most importantly, the ED and TD can cancel the scattering of each other when $\textbf{P}=-ik\textbf{T}$. This corresponds exactly to the elusive non-trivial non-radiating charge-current distribution first proposed in 1995~\cite{Afanasiev1995}, also known as {\em anapole}~\cite{miroshnichenko2014seeing}. The challenge to satisfy this condition is that in most structures the TD moment is negligible:  $|ik\textbf{T}|\ll |\textbf{P}|$. Only recently such long sought non-radiation source has been achieved in carefully designed structures~\cite{Fedotov2013_SR,Basharin2014_arXiv,miroshnichenko2014seeing}.

\section{Interferences between toroidal and electric dipoles in high permittivity dielectric spheres}

In Fig.~\rpict{fig1}(a) the schematic of the scattering problem is shown: an incident plane wave is scattered by a spherical nanoparticle, which could be homogenous or multi-layered. The plane wave is polarized along $x$ direction (in terms of electric field) and propagating along $z$ direction. Figure~\rpict{fig1}(b) gives a schematic illustration of the magnetic field and current distribution of the TD excited in the spherical particle. As is shown, the TD can also be interpreted as a series of magnetic dipoles (enclosed circulating currents) aligned along a circle. The circle actually also represents the field-lines of the induced magnetic field \textbf{H}.

According to Mie theory, the total scattered power by the spherical particle can be expressed as (with no magnetic materials and thus relative permeability $\mu=1$)~\cite{Bohren1983_book}:

\begin{equation}
\label{all}
{\rm W_{\rm all}} ={{\pi \left| {E_0 } \right|^2 } \over {k\omega \mu _0 }}\sum\limits_{m = 1}^\infty  {(2m + 1)({{\left| {{a_m}} \right|}^2}}  + {\left| {{b_m}} \right|^2}),
\end{equation}
where $E_0$ is the magnitude of the incident electric field and $\mu_0$ is the permeability in vacuum. It is clear that the contribution associated with $\textbf{P}(a_1)$ is: $\rm W_{\textbf{P}(a_1)}=
{{3\pi \left| {E_0 a_1 } \right|^2 } \mathord{\left/{\vphantom {{3\pi \left| {E_0 a_1} \right|^2 } {k\omega \mu_0}}} \right.\kern-\nulldelimiterspace} {k\omega \mu _0 }}$. Meanwhile, the scattered power of Cartesian ED and TD are, respectively:
\begin{equation}
\label{ED_TD_power}
\rm W_{\rm \textbf{P}}  = {{\mu _0 \omega ^4 } \over {12\pi c}}\left| \textbf{P} \right|^2, ~~\rm W_{\rm \textbf{T}}  = {{\mu _0 \omega ^4 k^2 } \over {12\pi c}}\left| \textbf{T}\right|^2.
\end{equation}

Recently high permittivity dielectric nanoparticles have attracted enormous attention due to the efficient excitation of magnetic dipoles in such structures in the optical regime~\cite{Kuznetsov2012_SciRep,Evlyukhin2012_NL,Liu2014_CPB}. The high permittivity can effectively reduce the relative wavelength inside the nanoparticle, and thus provide sufficiently large optical path to support enclosed circulating currents, which results in excitation of optically-induced magnetic dipoles.  Considering the current distribution of  the TD (a series of circulating currents) shown in Fig.~\rpict{fig1}(b), there should also a possibility for TD excitation in high-index dielectric nanoparticles in the optical spectral regime.  As a first step, we study the scattering of a high permittivity dielectric sphere (refractive index $n=3.5$, radius $R=100$~nm).

The scattered power spectra of such a dielectric sphere is shown in Fig.~\rpict{fig2}(a). Besides the total scattered power (all, green curve), we have also shown the contributions from the far-field deduced dipolar [$\rm W_{\textbf{P}(a_1)}$], ED ($\rm W_{\textbf{P}}$) and TD ($\rm W_{\textbf{T}}$) modes. It is clear from Fig.~\rpict{fig2}(a) that: (1) The excitation efficiency of the TD would be higher at shorter wavelengths, and at long wavelength the excitation is negligible. This justifies the dropping of the TD moment term in conventional Cartesian multipole expansion method under long wavelength approximations; (2) At short wavelength with TD moment comparable to that of ED, the spectra of far-field deduced dipole [$\textbf{P}(a_1)$] is contrastingly different from that of the ED obtained through current integration ($\textbf{P}$). This is due to the fact that the deduction of $\textbf{P}(a_1)$ has combined the contributions of both ED and TD~\cite{Bohren1983_book,Doyle1989_optical,Wheeler2006_PRB} [see Eq.(\ref{Scattering})].  As it was demonstrated in Ref.~\cite{miroshnichenko2014seeing} $\textbf{P}(a_1)$ can be viewed as a superposition of $\textbf{P}$ and $\textbf{T}$: $\textbf{P}(a_1)\approx\textbf{P}+ik\textbf{T}$.  But at the same time we note that the above relation does not strictly hold, with the discrepancies from the higher order terms of Taylor expansions (with respect to $kr$) for the spherical Bessel functions (the electric, magnetic and toroidal multipoles correspond to the fundamental expansion terms of the zeroth order)~\cite{Radescu2002_PRE}. This is clear in Fig.~\rpict{fig2}(a) in the regime of approximately $530$~nm-$600$~nm, when even though TD is negligible there are still significant differences between $\textbf{P}(a_1)$ and $\textbf{P}$. However, at longer wavelengths the spectrum of $\textbf{P}(a_1)$ agrees well with that of $\textbf{P}$, as under long wavelength approximation the toroidal moments and higher order terms of Taylor expansions  of the spherical Bessel functions are negligible~\cite{Radescu2002_PRE,Fedotov2013_SR,miroshnichenko2014seeing}.

To give more details of the multipole excited, we select three points [marked by A, B, C in  Fig.~\rpict{fig2}(a)] and show the corresponding near-field distributions (on the $x$-$z$ plane with $y=0$) in Fig.~\rpict{fig2}(a-A)-Fig.~\rpict{fig2}(a-C), where the color-plots correspond to the perpendicular magnetic fields along $y$ ($\textbf{H}_y$) and the vector-plots correspond to the electric fields [or the current flow according to Eq.(\ref{current})] on plane. We should emphasize that in the spectral regime investigated in Fig.~\rpict{fig2}(a), despite the existence of ED and TD, there are also effective excitations of the magnetic dipole and quadrupole (see the right and left peaks of the green curve which correspond to their central resonant positions, respectively) and other higher order modes~\cite{Liu2012_ACSNANO,Liu2014_CPB,Liu2014_ultradirectional}. To show clearly the near fields of sole ED and TD, in Fig.~\rpict{fig2}(a-A)-Fig.~\rpict{fig2}(a-C) and Fig.~\rpict{fig2}(b-A) we have only included transverse-magnetic components (no radial magnetic field components) of the first order. All transverse-electric fields [no radial electric field components, which do not contribute to the integration in Eq.(\ref{Scattering})] and transverse-magnetic modes of higher orders have been dropped~\cite{Bohren1983_book} (see the appendix for more information).

 For the three points indicated in Fig.~\rpict{fig2}(a), we note that: (1) At point A ($\lambda$=$700$~nm), the TD excitation can be neglected and the near-field distribution [Fig.~\rpict{fig2}(a-A)] exhibits a typical ED response, with no loop currents inside the nanoparticle; (2) At point B ($\lambda$=$490$~nm), the ED is negligible compared to the TD, and in Fig.~\rpict{fig2}(a-B) there are two oppositely circulating current loops with the magnetic field confined within the loop. This corresponds exactly to the field-current distribution of a typical TD shown in Fig.~\rpict{fig1}(b), thus reconfirming the effective excitation of TD at this point; (3) At point C ($\lambda$=$462$~nm), there are both TD and ED excitation and more importantly they have the same amount of scattering power, which corresponds to the condition of $\textbf{P}=-ik\textbf{T}$, thus leading to the cancellation of dipolar scattering [see Eq.(\ref{Scattering}) and the black cure of $\textbf{P}(a_1)$ in Fig.~\rpict{fig2}(a)]. This can be further confirmed by the near-field distribution shown in Fig.~\rpict{fig2}(a-C), where the fields are almost zero outside the particle.  This is in sharp difference to the situation at point B [Fig.~\rpict{fig2}(a-B)]: since there is no ED excitation to cancel the scattering of the TD, and consequently there are still significant fields outside the sphere. We note here that the near-field distribution at the cancellation point B is similar to that of the Lamb-Chaplygin dipole vortex in fluid dynamics~\cite{Meleshko1994_on}.

For comparison, we have also shown the scattered power spectra in Fig.~\rpict{fig2}(b) for a silver sphere of radius $R=80$~nm. For the permittivity of silver we adopt the experimental data from Ref.~\cite{Johnson1972_PRB}. It is clear that in the whole spectrum regime, the magnitude of TD moment is far smaller than that of ED, and thus their interferences can be neglected according to Eq.(\ref{Scattering}).  The near-field distribution of point A marked in Fig.~\rpict{fig2}(b) at $\lambda=450$~nm is shown in Fig.~\rpict{fig2}(b-A) (the vector-plot has been confined within the particle for better vision), where a typical ED field distribution is exhibited.

\pict[0.7]{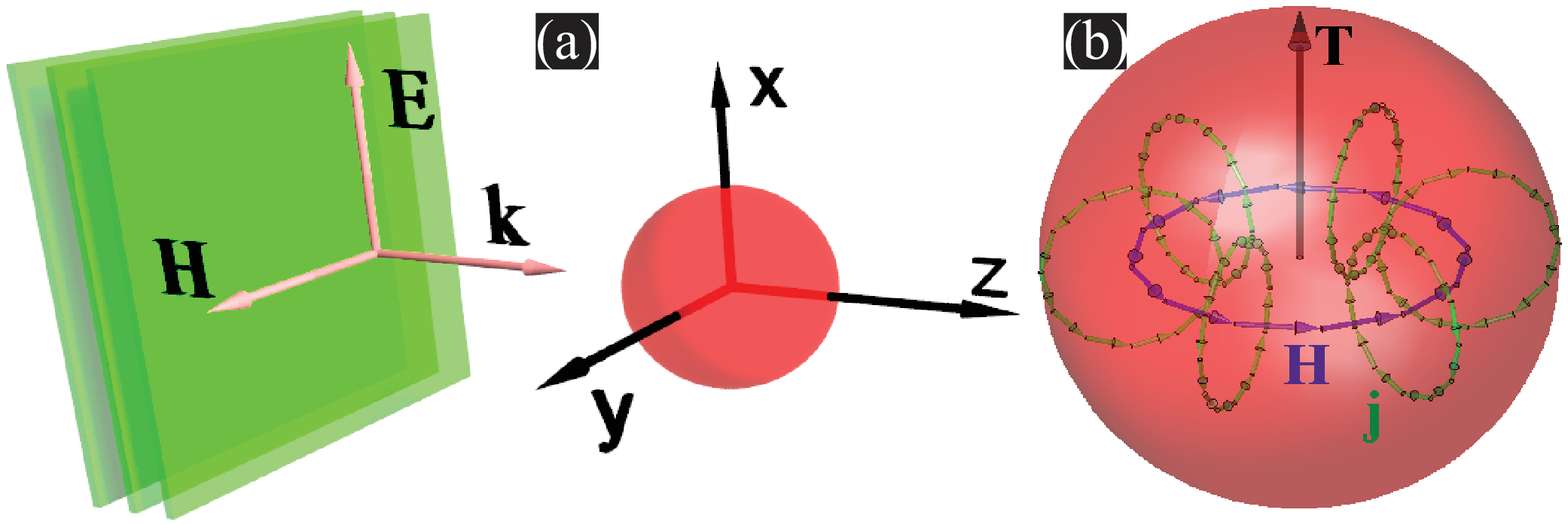}{fig1}{\small (Color online)  (a) The scattering of a spherical particle. The incident wave is a plane wave,  which is propagating along $z$ direction and polarized (in terms  of electric field) along $x$ direction. (b) Schematic illustration of the toroidal dipole excitation within the spherical particle. Both the current [or $\textbf{E}$ field according to Eq.(\ref{current})] and magnetic field $\textbf{H}$ distributions have been shown.}

\pict[0.99]{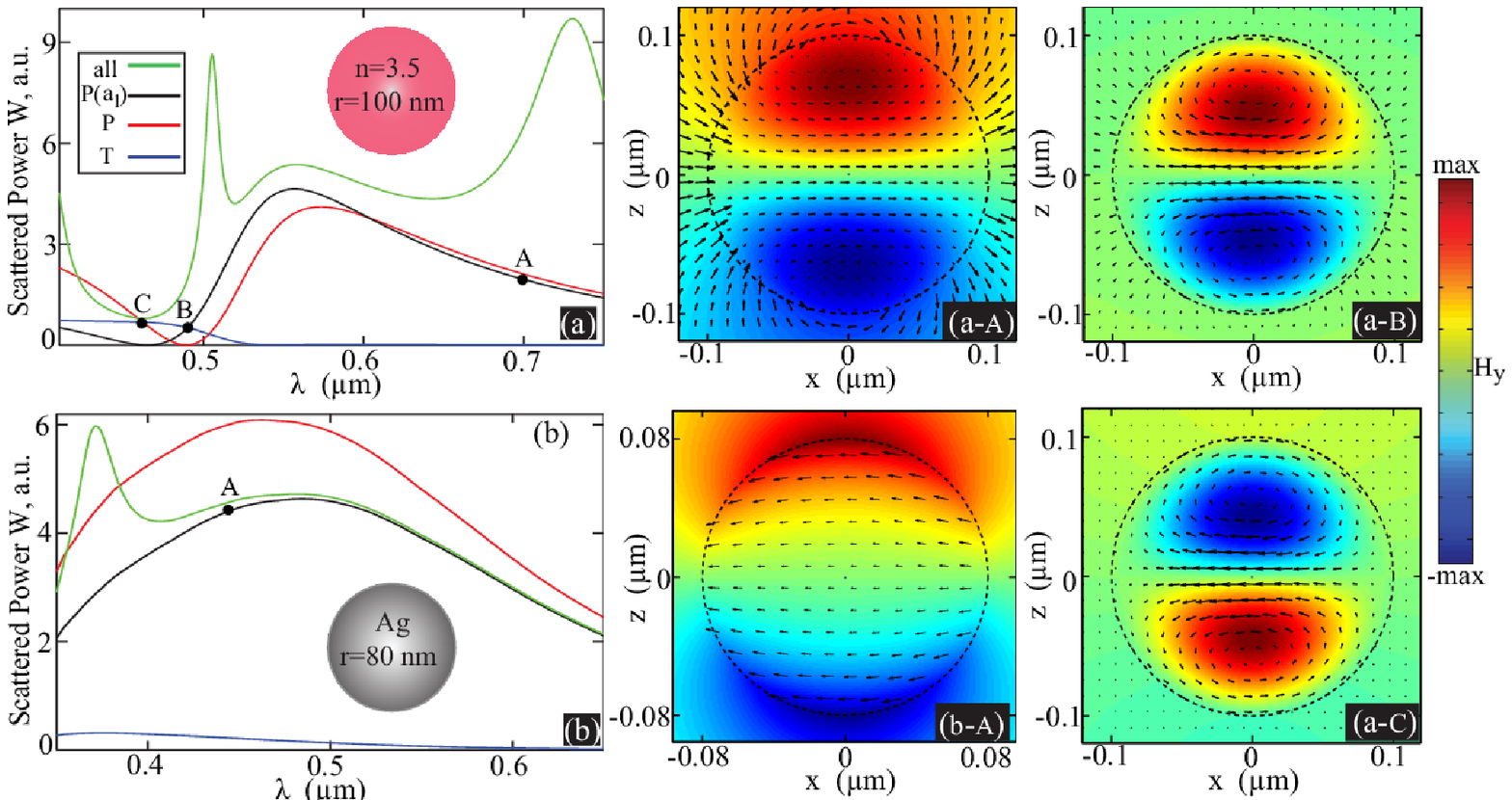}{fig2}{\small (Color online) Scattered power spectra for (a) a dielectric ($n=3.5$) sphere (inset) of $R=100$~nm and (b) a silver sphere (inset) of $R=80$~nm. For both cases, the total scattered power (all, green curves), and the contributions from far-field deduced dipole [$\rm W_{\textbf{P}(a_1)}$, black curves], electric dipole ($\rm W_{\textbf{P}}$, red curves) and toroidal dipole ($\rm W_{\textbf{T}}$, blue curves) are shown. Three points are indicated in (a) (A: $\lambda$=$700$~nm; B: $\lambda$=$490$~nm; C: $\lambda$=$462$~nm) and the corresponding near fields on the $x-z$ plane are shown respectively in (a-A)-(a-C).  The distributions for both $\textbf{H}_y$ (color-plots) and $\textbf{E}$ (or currents, vector-plots) are shown. The point A marked in (b) corresponds to $\lambda=450$~nm and the near-field distribution is shown in (b-A).}

\section{Toroidal dipole induced non-trivial transparency for core-shell nanoparticles}

To obtain a non-trivial transparency window induced by the complete destructive interference of ED and TD as indicated by Eq.(\ref{Scattering}), the following two conditions have to be satisfied: (1) There are effective ED and TD excitation with $\textbf{P}(a_1)\approx\textbf{P}+ik\textbf{T}$; (2) The excitations of all other multipoles have been significantly suppressed.  At point C in Fig.~\rpict{fig2}(a), though there is a complete destructive interference between ED and TD, it is not really a transparency window as the second condition has not been satisfied [the total scattering is not negligible compared to the scattering of ED or TD, see the green curve in Fig.~\rpict{fig2}(a)]. The scattering at point C has mainly been contributed
by the magnetic dipole (MD) and magnetic quadrupole (MQ), and the electric quadrupole~\cite{Liu2014_ultradirectional} (EQ, see the appendix for the definitions of those multipoles and their scattered power).

It was recently demonstrated that in metal-dielectric core-shell nanoparticles the positions of electric multipoles can be tuned effectively without significantly changing the positions of magnetic multipoles~\cite{Paniague2011_NJP,Liu2012_ACSNANO}.  This means that in such nanoparticles, we have sufficient freedom to tune the cancellation point of ED and TD to the minimum scattering point of MD and MQ where they are significantly suppressed, thus achieving a non-trivial TD induced transparency.  To confirm this, in  Fig.~\rpict{fig3}(a) we show the scattered power spectra of a metal (Ag)-dielectric ($n=3.5$) core-shell nanosphere with inner radius $R_1=20$~nm and outer radius $R_2=213$~nm. Besides the multipoles shown in Fig.~\rpict{fig2}(a), we have also included the contributions from MD ($\rm W_{\rm \textbf{M}}$), MQ ($\rm W_{\rm \textbf{Q}_{\rm m}} $), and EQ ($\rm W_{\rm \textbf{Q}_{\rm e}}$) as indicated. The point A ($\lambda=792$~nm) marked corresponds to the complete destructive interference point of ED and TD. This can be further confirmed by the fact that at this point the magnitude of $\textbf{P}(a_1)$ is almost zero. Moreover, at this point all the other multipole moments have been significantly suppressed, with a negligible total scattering as indicated [see the green curve of in Fig.~\rpict{fig3}(a)].  In other words, we have obtained the non-trivial  transparency or a non-radiating source induced by the interferences of ED and TD within an individual scattering particle in the optical spectral regime. Although, such scattering suppression has been achieved, this condition corresponds to the dip of the toroidal moment,  meaning that the magnitudes of ED and TD are very small at this point, which unfortunately are actually very close to that of the other multipole moments. Moreover, it is not really a transparency point as the total scattering is not negligible compared to the scattering of ED and TD.  In our case however, we have achieved the transparency point close to the peak of the TD, where the ED and TD are much stronger than other multipoles the total scattering is negligible.

For comparison, we also investigate another dielectric ($n=3.5$)-Ag core-shell nanosphere with $R_1=50$~nm and radius $R_2=80$~nm, and show the  corresponding scattered power spectra in Fig.~\rpict{fig3}(b) (contributions from higher-order multipoles are far smaller and thus not shown). It is clear that there is also a transparency point (marked by point A at $\lambda=694$~nm). In sharp contrast to the point indicated in Fig.~\rpict{fig3}(a), at this point, all the multipoles including ED and TD are not effectively excited, thus leading to a trivial transparency window. To confirm further the differences between these two points indicated in  Fig.~\rpict{fig3}(a) and Fig.~\rpict{fig3}(b), we show the  corresponding near-field distributions in Fig.~\rpict{fig3}(a-A) and Fig.~\rpict{fig3}(b-A) respectively. Though for both cases, the incident plane waves have experienced almost no scattering (no perturbation of the wave-front), while for the former case there is effective mode excitation within the nanoparticle (non-trivial transparency) and for the latter one, there are almost null fields inside (trivial transparency).

We note here that though there have already been extensive studies on the transparency phenomena in core-shell nanoparticles (see Refs.~\cite{Kerker1975_JOSA,Chew1976_abnormally,Alu2005_PRE} and many citing articles that followed), the crucial roles that could be played by toroidal moments have not been identified so far. For the structures proposed by Kerker \textit{et al.} in Refs.~\cite{Kerker1975_JOSA,Chew1976_abnormally}, the transparency was induced by the annihilation of the ED (dipolar polarization) due to the contributions of opposite signs and of the same magnitude from different layers of the particles [see Eq.(\ref{current}) and Eq.(\ref{ED_TD})]. As the particles proposed are much smaller than the wavelength (in the electrostatic regime), the toroidal moments have not been excited. For the studies conducted by Alu and Engheta which are not confined to the electrostatic regime and thus more comprehensive~\cite{Alu2005_PRE}, though toroidal moments had been effectively excited, but their interpretations of the transparency are still the same as that of Kerker without recognising the contributions of toroidal moments, and thus are inaccurate for some of the structures they investigated. Only here in our work a more complete description of the transparencies of core-shell particles has been given.

\pict[0.8]{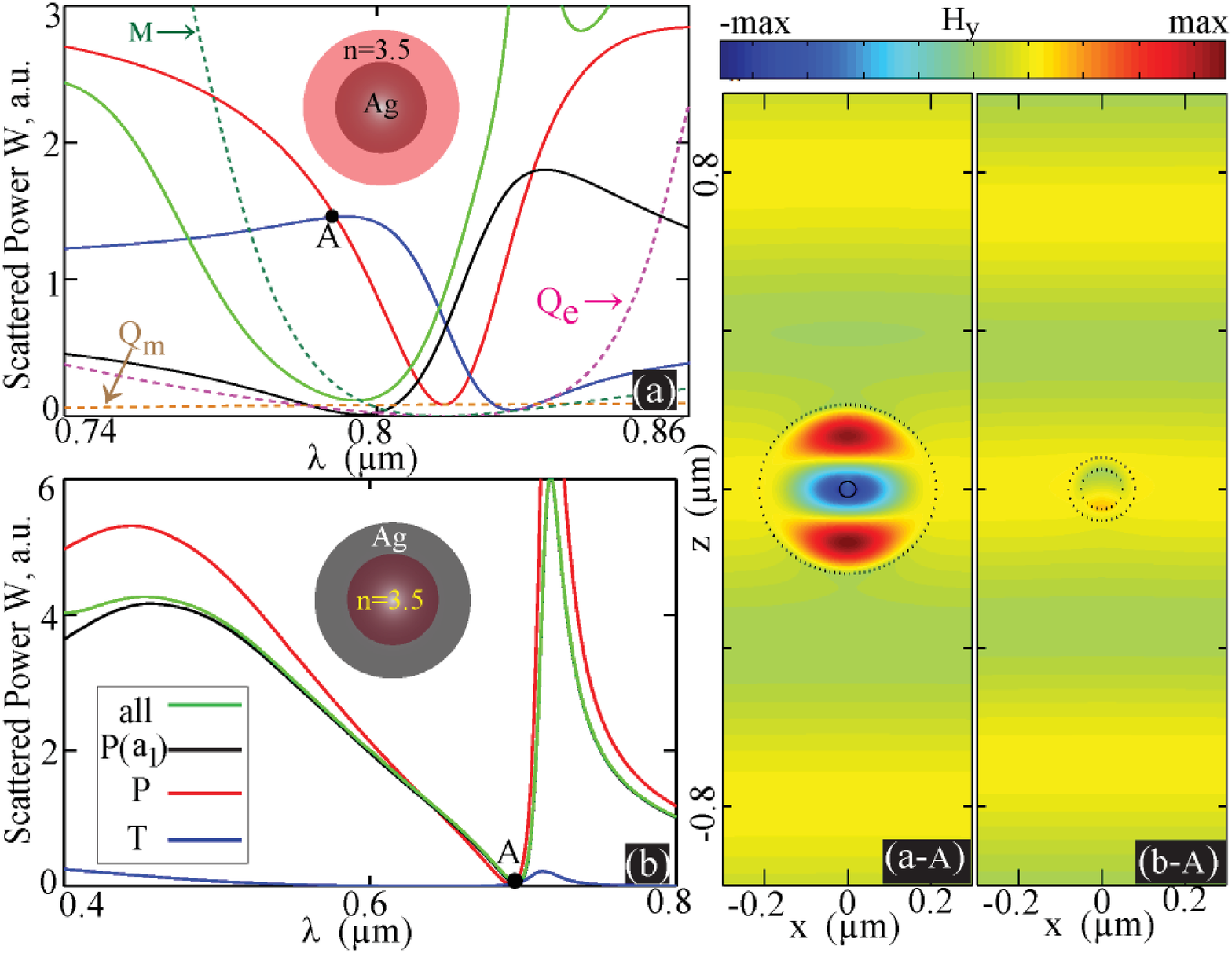}{fig3}{\small (Color online) Scattered power spectra for (a) a silver-dielectric ($n=3.5$) core-shell sphere (inset) of $R_1=20$~nm and $R_2=213$~nm, and (b) a dielectric ($n=3.5$)-silver core-shell sphere (inset) of $R_1=50$~nm and $R_2=80$~nm. In (a) the contributions from MD ($\rm W_{\rm \textbf{M}}$), MQ ($\rm W_{\rm\textbf{Q}_{\rm m}} $), and EQ ($\rm W_{\rm \textbf{Q}_{\rm e}}$) have also been included. For both cases transparency points have been indicated [$\lambda=792$~nm in (a) and $\lambda=694$~nm in (b)] and the corresponding near-field distributions are shown in (a-A) and (b-A) respectively.}

\section{Conclusions and Discussions}

By employing the complete Cartesian multipole expansion method incorporating toroidal multipoles, we revisited the Mie scattering problem of spherical nanoparticles. It is found that in high permittivity dielectric spheres toroidal dipoles can be effectively excited with comparable magnitude to that of the electric dipole, and thus can affect significantly the overall dipolar scattering. We further investigate the scattering transparencies of core-shell nanoparticles and reveal that they can be classified into two categories:  the trivial transparency without exciting the multipole moments effectively inside the particle, and the non-trivial transparency induced by the destructive interferences of electric and toroidal dipoles, which exactly corresponds to a non-trivial non-radiating source.

We have confined our study to the toroidal dipoles and the transparencies induced by their destructive interferences with electric dipoles. The investigations can certainly be extended to higher order toroidal modes and the induced transparencies originating from the interferences between higher order electric and toroidal modes. Such studies be conducted in larger particles and/or in smaller wavelength spectral regime, where there is sufficiently large optical path for loops of circulating currents.  Here we have discussed only spherical particles, in which the fields can be expressed analytically, and thus all the multipole moments can be calculated accurately. Similar studies can also be conducted for other non-spherical structures where the fields and currents, and thus multipole moments can be obtained through numerical methods~\cite{miroshnichenko2014seeing}.

Recently it is shown that the interferences of electric and magnetic multipoles provides a lot of extra freedom for scattering pattern shaping of scattering particles (see Ref. \cite{Liu2014_CPB} and references therein). We anticipate that when toroidal multipoles are considered, more flexibilities will be obtained for the scattering manipulation. Considering that the toroidal multipoles are highly related to both the local near-field distributions and the far-field scattering patterns, the introduction of toroidal moments into scattering particles may shed new light to many related studies and can possibly expand the applications of scattering particles in fields including sensing, nanoantennas, and photovoltaic devices. Moreover, investigations into electromagnetic light-matter interactions involving toroidal moments can possibly merge with the emerging fields of topological photonics and low-dimensional photonics, and help to establish new platforms for the study of many exotic optical effects, such as the dynamical Aharonov-Bohm effect and its possible applications in quantum information~\cite{Afanasiev1995,Marengo2002_Nonradiating}.

\section{Acknowledgements}

 We thank B. Lei, H. Hu, D. N. Neshev, R. F. Oulton  and Y. S. Kivshar for many fruitful discussions, and acknowledge the financial support from the National Natural Science Foundation of China (Grant numbers: $11404403$ and $11304389$), the Australian Research Council (FT110100037) and the Basic Research Scheme of College of Optoelectronic Science and Engineering, National University of Defence Technology.

\appendix

\section{Magnetic dipole moment, higher order multipoles  and their scattered power}

Similar to the definitions of ED and TD in Eq.~(\ref{ED_TD}), based on the current distribution [Eq.~(\ref{current})]  the magnetic dipole moment ($\textbf{M}$), magnetic quadrupole moment ($\textbf{Q}_{\rm m}$), and electric quadrupole moment ($\textbf{Q}_{\rm e}$) can be expressed in Cartesian basis ($\alpha,\beta=x,y,z$) as follows~\cite{Radescu2002_PRE,Kaelberer2010_Science,Fedotov2013_SR}:

\begin{eqnarray}
\label{other_moments}
&\textbf{M}^{\alpha} = {1 \over {2c}}\int {d^3 r\left[ {\textbf{r }\times \textbf{J}(\textbf{r})} \right]^{]\alpha}},\\
&\textbf{Q}_{\rm m}^{\alpha,\beta}  = {1 \over {3c}}\int {d^3 r\left[ {(\textbf{r} \times \textbf{J}(\textbf{r})) \otimes \textbf{r} + \textbf{r} \otimes (\textbf{r} \times \textbf{J}(\textbf{r}))} \right]^{\alpha,\beta}} ,\\
&\textbf{Q}_{\rm e}^{\alpha,\beta}  = {1 \over {-iw}}\int {d^3 r\left[ {(\textbf{r }\otimes \textbf{J}(\textbf{r}) + \textbf{J}(\textbf{r}) \otimes \textbf{r})^{\alpha,\beta}  - {2 \over 3}\delta _{\alpha,\beta} \textbf{r} \cdot \textbf{J}(\textbf{r})} \right]},
\end{eqnarray}

where $\otimes$ denotes the tensor product.  The corresponding scattered power is~\cite{Radescu2002_PRE,Kaelberer2010_Science,Fedotov2013_SR}:
\begin{eqnarray}
\label{other_moments_power}
&\rm W_{\rm \textbf{M}}  = {{\mu _0 \omega ^4 } \over {12\pi c}}\left| \textbf{M} \right|^2,\\
&\rm W_{\rm \textbf{Q}_{\rm m}}  = {{\mu _0 \omega ^4 k^2} \over {160\pi c}}\left| {\textbf{Q}_{\rm m}} \right|^2={{\mu _0 \omega ^4 k^2} \over {160\pi c}}\sum_{\alpha,\beta}\left| {\textbf{Q}^{\alpha,\beta}_{\rm m}} \right|^2,\\
&\rm W_{\rm\textbf{Q}_{\rm e}}  = {{\mu _0 \omega ^4 k^2} \over {160\pi c}}\left| {\textbf{Q}_{\rm e}} \right|^2={{\mu _0 \omega ^4 k^2} \over {160\pi c}}\sum_{\alpha,\beta}\left| {\textbf{Q}^{\alpha,\beta}_{\rm e}} \right|^2.
\end{eqnarray}

\section{Scattered field decompositions in spherical vector harmonics}

For an incident plane wave, the scattered fields of a spherical particle can be expressed as~\cite{Bohren1983_book}:

\begin{eqnarray}
&\textbf{E}_{\rm s}  = \sum\limits_{m = 1}^\infty  {E_m } [ia_m \textbf{N}_{e,m} (\textbf{r}) - b_m \textbf{M}_{o,m} (\textbf{r})],\label{scattered_fields1}\\
&\textbf{H}_{\rm s}  = {k \over {\omega \mu _0 }}\sum\limits_{m = 1}^\infty  {E_m } [ib_m \textbf{N}_{o,m} (\textbf{r}) + a_m \textbf{M}_{e,m} (\textbf{r})]\label{scattered_fields2},
\end{eqnarray}
where $E_m  = i^m E_0 {{2m + 1} \over {m(m + 1)}}$ (magnitude of the electric field of the incident field is $E_{\rm 0}$), and the expressions for $\textbf{M}$ and $\textbf{N}$ are tedious, which could be found in Ref~\cite{Bohren1983_book}. It is worth noting that the field components expressed in $\textbf{M}$ are transverse with respect to $\textbf{r}$ ($\textbf{M} \cdot \textbf{r} = 0$). According to Eqs.~(\ref{scattered_fields1}-\ref{scattered_fields2}), this indicates that the scattered fields shown above have actually been decomposed into transverse-magnetic (TM) parts (associated with $a_{\rm m}$) and transverse-electric (TE) parts (associated with $b_{\rm m}$). This explains why $a_{\rm m}$ and  $b_{\rm m}$ have been revealed (through comparing far-field scattering patterns) to be directly related respectively to the magnitudes of electric and  magnetic moments excited within the spherical particles~\cite{Doyle1989_optical,Wheeler2006_PRB}

Similar to the scattered fields shown in Eqs.~(\ref{scattered_fields1}-\ref{scattered_fields2}), the fields inside the spherical particles (single-layered or multi-layered) can be also decomposed into TM and TE components in different orders.  For the near-field plots in Fig.~\rpict{fig2}(a-A)-Fig.~\rpict{fig2}(a-C) and Fig.~\rpict{fig2}(b-A), only the TM components of the first order have been included. All the TE components and TM components of higher orders have not been included for clearer illustration of the ED and TD. In Fig.~\rpict{fig3} however all the field components have been included and the excitations of other multipoles are negligible.

\bibliography{References_scattering_Anapole2}
\end{document}